\documentclass[a4paper]{article}
\usepackage{graphicx}
\usepackage{a4wide}
\usepackage[affil-it,auth-sc]{authblk}

\newcommand{\MGF}{\vec{\cal B}}
\newcommand{\AVP}{\vec{\cal A}}
\newcommand{\rmd}{{\rm d}}
\newcommand{\rme}{{\rm e}}
\newcommand{\rmi}{{\rm i}}
\newcommand{\mbfj}{{\mathbf{j}}}
\newcommand{\mbfk}{{\mathbf{k}}}
\newcommand{\mbfo}{{\mathbf{o}}}
\newcommand{\mbfp}{{\mathbf{p}}}
\newcommand{\mbfr}{{\mathbf{r}}}

\newcommand{\BesselI}{{\rm I}}
\newcommand{\BesselJ}{{\rm J}}

\newcommand{\Ai}{{\rm Ai}}

\newcommand{\bra}{\langle}
\newcommand{\ket}{\rangle}

\begin{document}
\title{An efficient and accurate method to obtain the energy-dependent Green function for general potentials}

\author[1,2]{Tobias Kramer\footnote{Homepage: http://www.quantumdynamics.de}}
\author[1,3]{Eric J Heller}
\author[1,4]{Robert E Parrott}
\affil[1]{Harvard University, Department of Physics, Cambridge, MA~02138, USA}
\affil[2]{Institute for Theoretical Physics, University of Regensburg, Universit\"atsstr.\ 31, 93053 Regensburg, Germany}
\affil[3]{Harvard University, Department of Chemistry and Chemical Biology, Cambridge, MA~02138, USA}
\affil[4]{Harvard University, School of Engineering and Applied Science, Cambridge, MA~02138, USA}

\maketitle

\begin{abstract}
Time-dependent quantum mechanics provides an intuitive picture of particle propagation in external fields. Semiclassical methods link the classical trajectories of particles with their quantum mechanical propagation.
Many analytical results  and a variety of numerical methods have been developed to solve the time-dependent Schr\"odinger equation. The time-dependent methods work for nearly arbitrarily shaped potentials, including  sources and sinks via complex-valued potentials. 
Many quantities are measured at fixed energy, which is seemingly not well suited  for a time-dependent formulation. Very few methods exist to obtain the energy-dependent Green function for complicated potentials without resorting to ensemble averages or using certain lead-in arrangements. Here, we demonstrate in detail a time-dependent approach, which can accurately and effectively construct the energy-dependent Green function for very general potentials. The applications of the method are numerous, including chemical, mesoscopic, and atomic physics.
\end{abstract}

\section{Introduction}

In principle, the energy Green function $G(\mbfr,\mbfr';E)$ contains  all one ever needs to know about a stationary system. For example, in atomic systems, the Green function gives the photoionization and photodetachment rate in the presence of electric and magnetic fields \cite{Kramer2002a,Bracher2005b}. The theory of atom lasers emitted from Bose-Einstein condensates relies on the Green function \cite{Kramer2006b}. The Green function is of course  quite nontrivial to obtain, even for single particle problems, if disorder, magnetic fields, etc.\ are present. 

In mesoscopic systems (which are characterized by large free path lengths) the Green function connects the transport of electrons through the system with the conductivity, and makes possible the determination of the microscopic electron flow. With the advent of imaging techniques which   spatially resolve the electron flow on the nano scale, the realistic modelling of a device, including charged gates, disorder potentials and screening, has become a necessity \cite{Topinka2001a}.

Traditionally, Green functions for  mesoscopic systems demanded either a diagrammatic approach based on perturbation theory (chap.~5 of \cite{Datta1997a}) or to connect leads to the scattering region to obtain the Green function of the full system consisting of leads and scattering region by recursion (chap.~3 of \cite{Ferry1997a}).
The diagrammatic approach has the serious drawback that it is limited to the ensemble average of Green functions for many realizations of the potential. The ensemble averaged current through a device is in general very different from the flow through a typical realization of the device.  The Green function for a single realization of the device, instead of an ensemble, has become central to many experiments.
The recursive Green function technique has serious limitations due to the attachment of waveguide-type leads to the system which have to be arranged colinearly with the scattering region (the last condition can be relaxed by employing a smart reordering scheme based on graph theory \cite{Wimmer2008a}).

The approach presented in this paper permits us to obtain the Green function for a single realization with high accuracy and short computational time -- without the introduction of leads. Our approach works for very general electron-sources attached to a device, i.e.\ the injection of current  to the two-dimensional electronic subsystem from a (3D) reservoir through a tunneling contact (somewhat similar to the current transport from the tip of a scanning tunneling microscope \cite{Donner2004a}). Also inner contacts (which are used i.e.\ in Mach-Zehnder interferometers in mesoscopic systems) are possible to incorporate in the time-dependent method, as well as spin-orbit interactions.

The Green function  contains the discrete eigenenergies $E_n$ and the continuous ones $E_\nu$, and all eigenfunctions $\psi_{n,\nu}$ of the Hamiltonian in a compact form:
\begin{equation}
G(\mbfr,\mbfr';E)=
\sum_{n}\frac{\psi_n(\mbfr)\psi_n(\mbfr')}{E-E_n}
+\int\rmd\nu\frac{\psi_\nu(\mbfr)\psi_\nu(\mbfr')}{E-E_\nu}.
\end{equation}
Surprisingly, only very few energy-dependent Green functions are known analytically, in contrast to a vast amount of available time-dependent propagators \cite{Grosche1998a}. The direct calculation of the Green function for a given potential is generally not possible and perturbation expansions diverge. Thus the spectral properties of many systems are unknown, even though they are very relevant for predicting and analyzing experimental results.

One way of obtaining an energy Green function is via Fourier transform from the time domain\cite{Tomsovic1991a,Tomsovic1993a,Heller1991b,Heller1978a,Heller1978b}.  This can be done numerically, as it is always a one dimensional   integral no matter how many degrees of freedom are present. However this method  of obtaining the Green function has not been fully exploited or explored. In this paper, we describe an approach for obtaining energy-dependent Green functions via   the time-dependent Schr\"odinger equation. In Sect.~\ref{sec:tdse} we establish the connection between the time-dependent Schr\"odinger equation, the autocorrelation function of a wavepacket, and the energy-dependent Green function. Analytical and numerical examples of the method are discussed in Sects.~\ref{sec:example} and \ref{sec:dfft}. We provide a quantitative analysis of the accuracy of the method in Sect.~\ref{sec:acc}. We apply our method to electron propagation through a nano device in Sect.~\ref{sec:application}. In Sect.~\ref{sec:temp} we discuss thermal wavepackets and finally give some conclusions.

\section{The time-dependent Schr\"odinger equation.}\label{sec:tdse}

In the time-dependent formulation of quantum mechanics, the use of the propagator $K$ permits us to transform the partial differential time-dependent Schr\"odinger equation for an initial quantum state $\psi(\mbfr',t_i)$ into an integral equation:
\begin{equation}\label{eq:psiprop}
\psi(\mbfr,t_f)=\int\rmd\mbfr'\;K(\mbfr,t_f|\mbfr',t_i)\psi(\mbfr',t_i)
\end{equation}
For a time-independent Hamiltonian, we can set $t=t_f-t_i$. The energy-dependent Green function is tied to the time-dependent propagator $K$ via a Laplace transform:
\begin{equation}\label{eq:GEnergyIntegral}
G(\mbfr,\mbfr';E)
=\frac{1}{\rmi\hbar}\lim_{\eta\rightarrow 0_+}
\int_0^\infty \rmd t\;\rme^{\rmi (E+\rmi\eta) t/\hbar}\;K(\mbfr,t|\mbfr',0),
\end{equation}
where $\eta$ is a small positive number, which selects the outgoing wave boundary condition at $\mbfr$ and thus the retarded Green function. However, the Laplace transform is seldomly used to obtain the Green function analytically or numerically (even if the propagator is known), since the propagator is a highly oscillatory function with additional singularities.

For Hamiltonians which are polynomials of positions and momenta up to quadratic order, the time-dependent propagator assumes the form
\begin{equation}
K(\mbfr,t|\mbfr',t')=a(t)\exp[\rmi S(\mbfr,t|\mbfr',t')/\hbar],
\end{equation}
where $S$ denotes the classical action, which is itself a maximally quadratic polynomial in positions. Therefore it is possible to integrate Eq.~(\ref{eq:psiprop}) analytically for Gaussian wavepackets in $D$ dimensions
since all integrals involve Gaussian functions. For non-quadratic Hamiltonians,
the analytic form of the propagator is in general not known. However, various methods have been developed to solve the time-dependent Schr\"odinger equation numerically (see Sec.~\ref{sec:dfft}).

\subsection{From wavepackets to the Green function}\label{sec:autocorr}

The energy-dependent Green function satisfies the inhomogeneous Schr{\"o}dinger equation
\begin{equation}
[H-E] G(\mbfr,\mbfr';E)=\delta(\mbfr-\mbfr')
\end{equation}
with a $\delta$-distribution ``source''-term. If we specify as source-term a spatially extended wavepacket $\psi(\mbfr)$, we obtain the overlap of the Green function with the state $\psi$
\begin{eqnarray}\label{eq:ldoswp}
\bra \psi | G(E) | \psi \ket
&=&\int \rmd\mbfr \int \rmd\mbfr' \psi(\mbfr,0)^* G(\mbfr,\mbfr';E) \psi(\mbfr',0)\nonumber\\
&=&\frac{1}{\rmi\hbar}\int_0^\infty\rmd t\;\rme^{\rmi E t/\hbar}
\int \rmd\mbfr \int \rmd\mbfr' \psi(\mbfr,0)^*K(\mbfr,t|\mbfr',0) \psi(\mbfr',0) \nonumber\\
&=&\frac{1}{\rmi\hbar}\int_0^\infty\rmd t\;\rme^{\rmi E t/\hbar}
\int \rmd\mbfr\; \psi(\mbfr,0)^* \psi(\mbfr,t) \nonumber\\
&=&\frac{1}{\rmi \hbar}\int_0^\infty\rmd t\;\rme^{\rmi E t/\hbar} C(t),\label{eq:autocorr_laplace}
\end{eqnarray}
where $C(t)$ denotes the autocorrelation function
\begin{equation}
C(t)=\int \rmd\mbfr\; \psi(\mbfr,0)^* \psi(\mbfr,t).
\end{equation}
The Green function for $\mbfr'\ne\mbfr$ is obtained by replacing $\bra\psi|$ with another state $\bra\tilde{\psi}_{\mbfr'}|$ centered around $\mbfr'$ while keeping $|\psi\ket$. In numerical calculations $\mbfr'$ is a grid point of the discretized representation of the potential and of the wave functions (see eq.~(\ref{eq:green})).

\subsection{Calculation of the local density of states (LDOS)}

The local density of states (LDOS) at point $\mbfr$ is the imaginary part of the Green function
\begin{equation}
n(\mbfr;E)=-\frac{1}{\pi}\Im G(\mbfr,\mbfr;E).
\end{equation}
The definition of the LDOS requires to use a $\delta$-distributed initial state, which cannot be represented by a finite width wavepacket. This is an important technical point, requiring some care to handle correctly. Essentially, a point source (as in the coordinate space Green function) implies very high (in fact infinite) energy components, which are inconvenient and physically unimportant in a practical sense. In Sect.~\ref{sec:acc} we discuss how to recover the Green function  from a finite width wavepacket propagation, to very high accuracy. 

\section{Analytical Examples: Free particle and motion in a uniform electric field}\label{sec:example}

In the following, we take as initial wavepacket in two dimensions a Gaussian of the form
\begin{equation}
\psi(\mbfr)=\frac{1}{a\sqrt{\pi}} 
\exp\left[-\frac{r^2}{2 a^2}\right].
\label{wp}
\end{equation}
The probability density of this wavepacket is normalized to unity. We obtain the LDOS in correct units, if we multiply the correlation function by
\begin{equation}
\underbrace{\frac{1}{4\pi a^2}}_{\delta}
{\frac{-1}{\pi\rmi\hbar}}=-\frac{1}{4\rmi\pi^2 a^2 \hbar},
\end{equation}
where the first contribution is the switch of the normalization from unity of the modulus squared $\int {|\psi(\mbfr)|}^2 \rmd^2\mbfr=1$ to the non-squared function $\int \psi(\mbfr) \rmd^2\mbfr=1$, (this in order to correspond to  the delta function normalization of a coordinate state $\vert {\bf r} \rangle $) and the second factor originates from getting the imaginary part by using 
Laplace transform .
Thus we obtain for the wave-packet-LDOS for a wavepacket centered around $\mbfr$
\begin{equation}
n^{(a)}(\mbfr;E)=-\frac{1}{4\pi^2 a^2} \Im \left[\frac{1}{\rmi\hbar} \int_{0}^\infty\rmd t \; \rme^{\rmi E t/\hbar} C(t) \right].
\end{equation}
which depends on parameters of the wavepacket (like the width $a$). For numerical reasons we do not want to be propagating wavepackets with very high energy components if we can avoid it.  In order to discover how to compensate for  the finite width of the wavepacket, we   calculate specific autocorrelation functions analytically for systems with continuous spectra.

\subsection{The free propagator in 2D}

For the Hamiltonian
\begin{equation}
H_{\rm free}=\frac{p_x^2+p_y^2}{2m}
\end{equation}
the propagator reads
\begin{equation}\label{eq:Kfree}
K_{\rm free}(\mbfr,t|\mbfr',0)
=\frac{m}{2\pi\rmi\hbar t}
\exp\left(\frac{\rmi m {|\mbfr-\mbfr'|}^2}{2\hbar t}\right).
\end{equation}
The four Gaussian integrations to obtain the autocorrelation function starting with the wavepacket in Eq.~(\ref{wp}) are straightforward and yield
\begin{equation}
C_{\rm free}(t)=\frac{2 a^2 m}{2 a^2 m+\rmi\hbar t}.
\end{equation}
The Laplace transform is given by
\begin{equation}
-\frac{\bra \psi |G(E)|\psi\ket}{4\pi^2 a^2}
=\frac{-1}{4\pi^2 a^2\rmi\hbar}\int_0^{\infty}\!\!\!\rmd t\,C_{\rm free}(t) \rme^{\rmi E t/\hbar}
=\frac{\rme^{-2a^2 m E/\hbar^2}m}{2\pi^2\hbar^2}
\left[2\pi\rmi+\Gamma(0,-2a^2 m E/\hbar^2)\right],
\end{equation}
where $\Gamma(a,x)$ denotes the incomplete Gamma function. Using $\Im\Gamma(0,-|x|)=-\pi$, we obtain the LDOS for the wavepacket centered around the origin $\mbfo$
\begin{equation}\label{eq:ldosafree}
n_{\rm free}^{(a)}(\mbfo;E)
=\Im\left[-\frac{\bra \psi |G(E)|\psi\ket}{4\pi^2 a^2} \right]
=\frac{m}{2\pi\hbar^2}\rme^{-2a^2 m E/\hbar^2}.
\end{equation}
The exponential decay of the wavepacket LDOS reflects the probability density of the kinetic energy of the initial wavepacket $\rho(E_{\rm kin})$, which suppresses higher energies exponentially.
In the momentum representation, with $k=\sqrt{k_x^2+k_y^2}$, the wavepacket becomes
\begin{equation}
\hat{\psi}(k,0)=\frac{1}{2\pi\hbar}\int\rmd\mbfr^2 \rme^{\rmi \mbfr\cdot\mbfk}\psi(\mbfr,0)=2a\sqrt{\pi}\rme^{-\frac{1}{2}a^2 k^2},
\end{equation}
and the energy probability distribution $\rho(E_{\rm kin})=2\pi m {|\hat{\psi}(k=\sqrt{2mE_{\rm kin}}/\hbar,0)|}^2/\hbar^2$ is properly normalized
\begin{equation}\label{eq:kinden}
\int_0^{\infty}\rho(E_{\rm kin})\rmd E_{\rm kin}=
2\pi\int_0^{\infty}\rmd k\,{|\hat{\psi}(k)|}^2=1.
\end{equation}
For $a=0$ we obtain the LDOS for a $\delta$-distribution.

\subsection{The linear potential in 2D}\label{sec:acefield}

For the linear potential with Hamiltonian and propagator
\begin{equation}
H_{\rm field}=\frac{p_x^2+p_y^2}{2m}-F x,\quad K_{\rm field}(\mbfr,t|\mbfr',0)
=\frac{m}{2\pi\rmi\hbar t}
\exp\left(\frac{\rmi m {|\mbfr-\mbfr'|}^2}{2\hbar t}+\frac{\rmi F t (x+x')}{2\hbar}-\frac{\rmi F^2t^3}{24\hbar m}\right),
\end{equation}
we obtain the following autocorrelation function
\begin{equation}
C_{\rm field}(t)=
\frac{2 a^2 m}{2 a^2 m+\rmi\hbar t}\exp\left({-\frac{a^2F^2t^2}{4\hbar^2}-\rmi\frac{F^2t^3}{24\hbar m}}\right).
\end{equation}
The LDOS for a Gaussian wavepacket of width $a$ centered around the origin $\mbfo$ becomes
\begin{equation}
n_{\rm field}^{(a)}(\mbfo;E)=\frac{m}{2\pi\hbar^2}\rme^{32\alpha^6/3-8 E \alpha^2 \beta}
\left[\frac{1}{3}-\Ai_1\left(2^{2/3}(4\alpha^4-2 E \beta)\right)\right]
\end{equation}
with
\begin{equation}
\beta={\left(\frac{m}{4\hbar^2 F^2}\right)}^{1/3},\quad 
\alpha=\beta F a,\quad
\Ai_1(z)=\int_0^z \rmd x\;\Ai(x)
\end{equation}
For $\alpha\rightarrow 0$ we recover the LDOS for a $\delta$-distribution
\begin{equation}
n_{\rm field}(\mbfo;E)=\frac{m}{2\pi\hbar^2}\left[\frac{1}{3}-
\Ai_1\left(2^{2/3}(-2 E \beta)\right)\right].
\end{equation}

\section{Numerical method for general potentials}\label{sec:dfft}

 The time evolution operator $U$ for a time-independent Hamiltonian $H$ with kinetic energy $T$ and potential energy $V$ can be divided into $N$ steps:
\begin{equation}
U(t,t')=\rme^{-\rmi H (t-t')/\hbar}
=\rme^{-\rmi (T+V) N\Delta t/\hbar}
={\left(\rme^{-\rmi (T+V) \Delta t/\hbar}\right)}^N
\end{equation}
where $\Delta t=(t-t')/N$. We separate the kinetic and potential energy terms
using the Baker-Campbell-Hausdorff series 
\begin{equation}
\rme^{-\rmi\Delta t/\hbar\;(V/2+T+V/2)}
=
\rme^{-\rmi\Delta t/\hbar\;V/2}
\rme^{-\rmi\Delta t/\hbar\;T}
\rme^{-\rmi\Delta t/\hbar\;V/2}
\rme^{{\cal O}(\Delta t^3)}.
\end{equation}
The symmetric distribution of $V/2$ cancels all terms of order $\Delta t^2$ and makes the approximation accurate up to order $\Delta t^3$ \cite{Feit1982a}. The application of the kinetic energy operator $\rme^{-\rmi\Delta t/\hbar\;T}$ to a wave function reduces to a multiplication if the wave function is given in momentum space, whereas the potential energy operator is a multiplication in position space. We denote the Fourier transforms between position and momentum space by ${\cal F}$:
\begin{equation}
{\cal F}\psi(\mbfr)=\frac{1}{{(2\pi\hbar)}^{D/2}}
\int\rmd\mbfr\;\rme^{-\rmi \mbfp\mbfr}\psi(\mbfr),
\quad
{\cal F}^{-1}\psi(\mbfp)=\frac{1}{{(2\pi\hbar)}^{D/2}}
\int\rmd\mbfp\;\rme^{\rmi \mbfp\mbfr}\psi(\mbfp).
\end{equation}
Combining $N$ consecutive steps and joining the half steps within the loop yields the solution of the time-dependent Schr\"odinger equation 
\begin{equation}
\psi(\mbfr,t'+N\Delta t)\approx
\rme^{-\rmi\Delta t/\hbar\;V/2}
{\left[
{\cal F}^{-1}
\rme^{-\rmi\Delta t/\hbar\;T}
{\cal F}
\rme^{-\rmi\Delta t/\hbar\;V}
\right]}^N
\rme^{\rmi\Delta t/\hbar\;V/2}\psi(\mbfr,t').
\end{equation}
Numerically, the wave function is represented on a grid and the value of $\psi$ on the grid points is used to perform a Discrete Fast Fourier Transform (DFFT).
Since we apply the numerical wavepacket method also to systems where the wavepacket will eventually leave the initial region of the numerical grid-representation, we have to either move the Fourier grid along with the center of the wavepacket or to remove the wavepacket at the borders of the region of interest. Otherwise the DFFT algorithm will let the wavepacket reappear at the other edge of the grid. The removal of the wavepacket in certain regions is achieved best by using a complex valued potential. By tracking the probability density, the absorbing potential region can actually simulate i.e.\ a contact (acting as an electron sink) and measure the electronic current taken out of the system in that region.

\section{Extracting the $\delta$-distribution LDOS from the autocorrelation function}\label{sec:acc}

\begin{figure}[t]
\begin{center}
\includegraphics[width=0.47\textwidth]{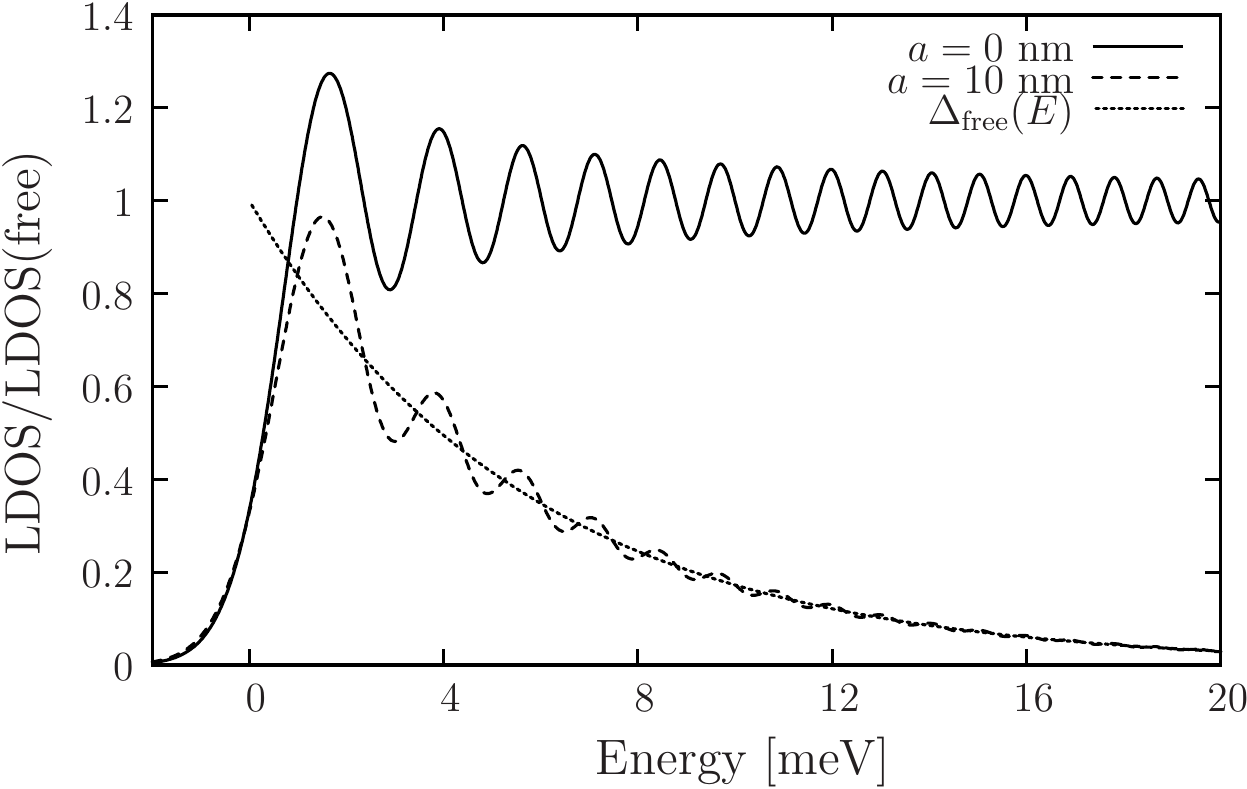}\hfill
\includegraphics[width=0.47\textwidth]{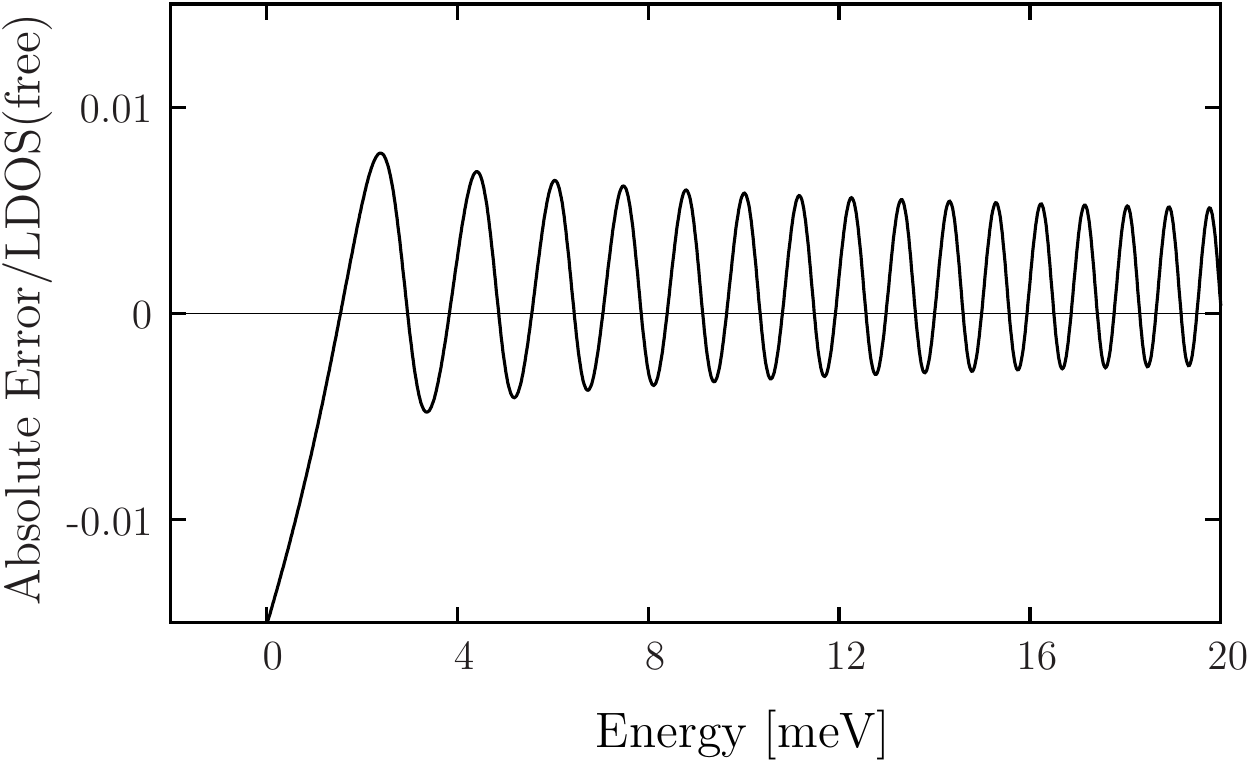}
\end{center}
\caption{Local density of states in a homogeneous electric field. Left panel: LDOS for a $\delta$-distribution ($a=0$), and for $a=10$~nm. Also shown is $\Delta_{\rm free}(E)$.
Right panel: Absolute error in units of the free-particle LDOS. Parameters: $F=50$~keV/m, effective mass $m=0.067\;m_e$.}
\label{fig:LDOSafield}
\end{figure}
With the examples above in mind we can see how to compensate for the energy cutoff  implied by a finite width wavepacket.
For a wavepacket with finite width $a$, the energy-dependent spectrum is exponentially damped for energies differing greatly from 
$E\approx\frac{\hbar^2}{2 m a^2}$. Choosing a smaller width $a$ is theoretically possible, but this would impose  serious limitations for the numerical propagation of wavepackets, demanding a finer grid of points in coordinate and smaller time steps. The wavepacket LDOS has to be corrected for   comparison with the $\delta$-distribution LDOS. The correction stems from the initial probability distribution of the kinetic energy. For the free-particle, the correction factor is given by Eq.~(\ref{eq:kinden})
\begin{equation}\label{eq:corrfac}
\Delta_{\rm free}(E)=\frac{n^{(a)}_{\rm free}(\mbfr;E)}{n_{\rm free}(\mbfr;E)}=\exp\left(-2a^2 m E/\hbar^2\right).
\end{equation}
The correction factor is dependent on the Hamiltonian under consideration, but $\Delta_{\rm free}(E)$ provides a lowest order estimate. In general, the correction factor is chosen to reproduce the kinetic energy-distribution of the initial wavepacket.

The analytic results of the previous section allow to analyze the error due to the use of the field-free correction factor for the linear electric field (see Fig.~\ref{fig:LDOSafield}). The LDOS for $a=0$ and $a=10$~nm are shown in the left panel. The right panel shows the absolute error of the corrected finite width calculation compared with the $a=0$ result. For the chosen values, the relative error is approximately 1/100.

\section{Application: Magnetic focusing}\label{sec:application}

\begin{figure}[t]
\begin{minipage}{0.495\textwidth}
\includegraphics[width=0.98\textwidth]{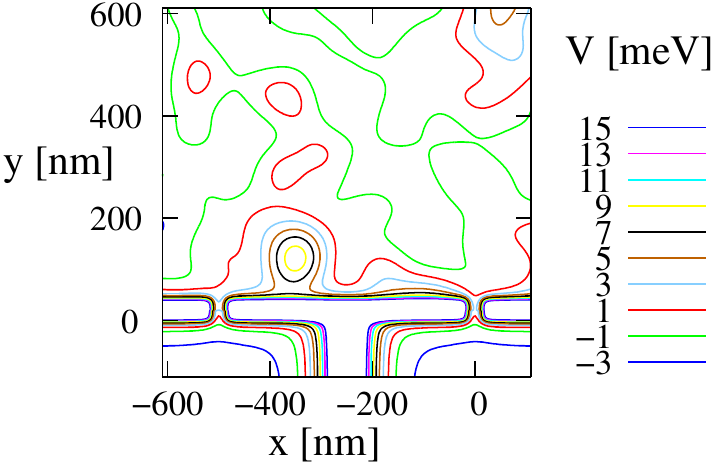}
\caption{Two-dimensional potential landscape for magnetic focusing. The T-shaped potential of two quantum point contacts forms two constrictions, while the background charges produce an irregular potential landscape.
}\label{fig:mf_pot}
\end{minipage}\hfill
\begin{minipage}{0.48\textwidth}
\includegraphics[width=0.99\textwidth]{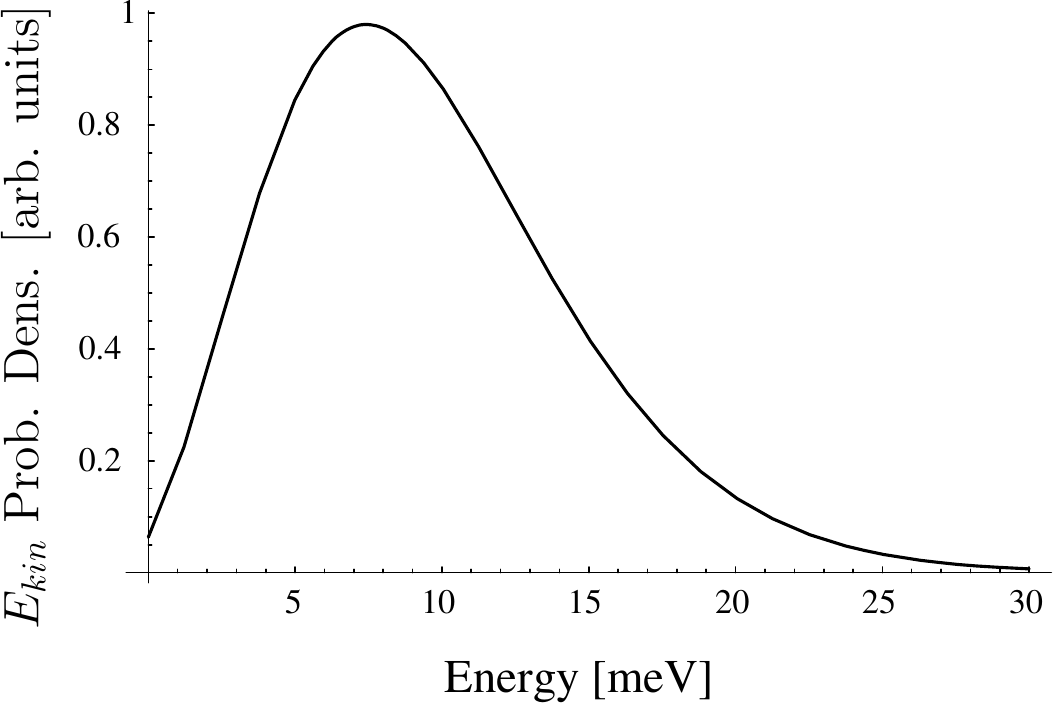}
\caption{Probability distribution of the kinetic energy in the initial wavepacket.
Parameters: $a=20$~nm, $k_0=\sqrt{2mE_0}/\hbar$, $E_0=9$~meV, effective mass $m=0.067 m_e$.}\label{fig:mf_wp}
\end{minipage} 
\end{figure}

The inclusion of magnetic fields necessitates splitting up the kinetic energy into two parts and to using a mixed momentum and position space representation. The use of a complex potential $V_{\rm abs}(\mbfr)$ is of special importance in the presence of magnetic fields, where periodic boundary conditions often fail to model the physical system under consideration. Magnetic focusing \cite{Aidala2007a} is such a case; here, periodic boundary conditions result in spurious edge currents around the chosen unit cell, which are absent in the real device due to additional contacts. The magnetic focusing setup consists of two separated constrictions (quantum point contacts) in a T-shaped two-dimensional device. The potential is shown in Fig.~\ref{fig:mf_pot}. The irregular potential landscape above the T-shaped potential produced by the gates is due to the presence of only partially screened background charges. It is important to model the effective potential landscape of the device realistically. Our model includes the effects of screening of gate and donor atom potentials and thus includes interactions on a mean field level \cite{Grill1995a,Stopa1996a,Davies1998a}. Experimentally, Ohmic contacts before the right and after the left constriction are connected to different voltages, which results in a current flowing through the system. The whole device is placed in a homogeneous magnetic field perpendicular to the two-dimensional plane.

The vector potential for a homogeneous magnetic field $\MGF=(0,0,{\cal B})$ in the symmetric gauge becomes $\AVP=(-y,x,0) {\cal B}/2$. The Hamiltonian splits into three parts
\begin{equation}
H=\underbrace{{\frac{p_x^2}{2m}-\omega_L p_x y}}_{T_{p_x,y}}
+\underbrace{\frac{p_y^2}{2m}+\omega_L p_y x}_{T_{p_y,x}}
+\underbrace{\frac{1}{2}\omega_L^2 (x^2+y^2)+V(x,y)}_{V{\rm eff}(x,y)},\quad
\omega_L=\frac{e {\cal B}}{2m},
\end{equation}
where the mixed momentum-position representation for the kinetic energy is possible since $[p_x,y]=[p_y,x]=0$. The new propagation algorithm becomes
\begin{eqnarray}
\psi(\mbfr,t'+N\Delta t)&\approx&
\rme^{-\rmi\Delta t/\hbar\;V_{\rm eff}/2}
{\left[
\rme^{-\rmi\Delta t/\hbar\;V_{\rm abs}}
{\cal F}^{-1}_y
\rme^{-\rmi\Delta t/\hbar\;T_{py,x}}
{\cal F}_y
{\cal F}^{-1}_x
\rme^{-\rmi\Delta t/\hbar\;T_{px,y}}
{\cal F}_x
\rme^{-\rmi\Delta t/\hbar\;V_{\rm eff}}
\right]}^N\nonumber\\
&&\quad\times\rme^{\rmi\Delta t/\hbar\;V_{\rm eff}/2}\psi(\mbfr,t'),
\end{eqnarray}
where ${\cal F}_x, {\cal F}_y$ denote partial Fourier transforms with respect to only one-dimension.

\begin{figure}[t]
\includegraphics[width=0.55\textwidth]{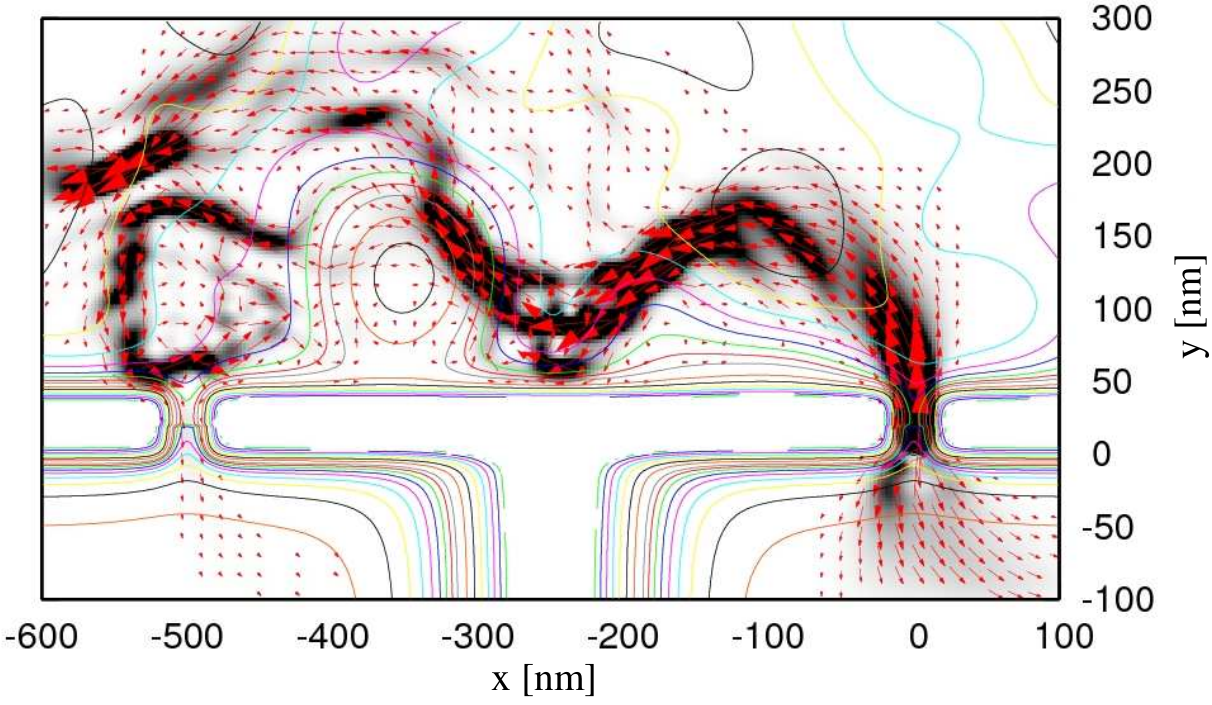}\hfill
\includegraphics[width=0.35\textwidth]{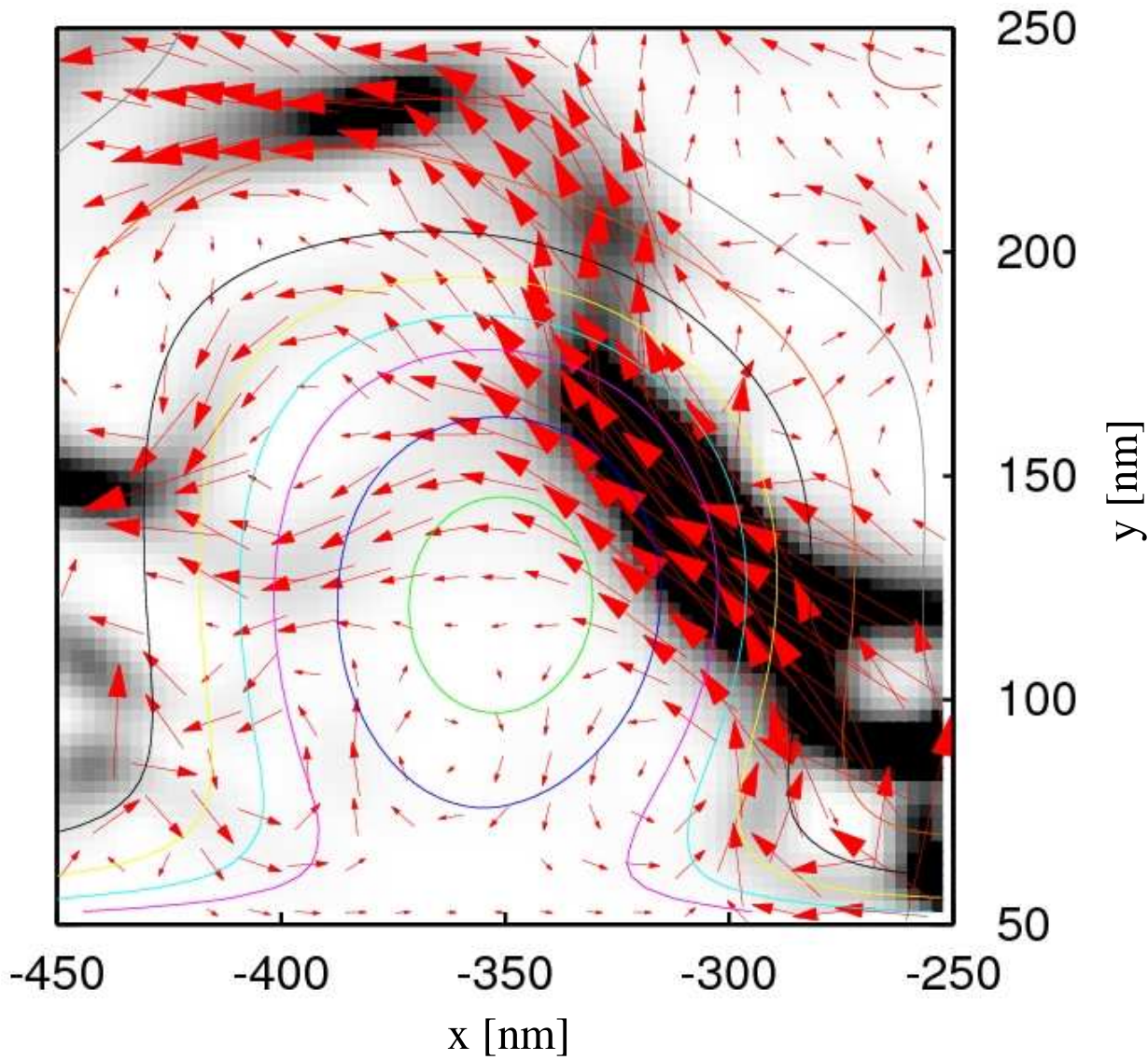}
\caption{Current flow through the device at $E=9$~meV. The wavepacket originates from $\mbfr'=(0,0)$. A magnetic field of ${\cal B}=0.8$~T is applied perpendicular to the $x-y$-plane. The dark regions denote an enhanced probability ${|G(\mbfr_{n,m},\mbfr';E)|}^2$, eq.~(\ref{eq:green}). 
The arrows denote the local current density, eq.~(\ref{eq:jden}). The right panel is a close-up of the current flow around an obstacle placed at $\mbfr=(-350,140)$~nm. Parameters: Effective mass $m=0.067\,m_e$.}\label{fig:mf_jdensity}
\end{figure}

In the following, we apply the wavepacket method in order to obtain a detailed picture of the current flow through the device. At $t=0$ we start a wavepacket which represents an expanding ``smoke ring'' centered around the origin :
\begin{equation}
\hat{\psi}(k_x,k_y,t=0)\propto\int_{-\pi}^\pi\frac{\rmd\theta}{2\pi}\exp\left(-a^2/2[{(k_x-k_0\cos\theta)}^2+{(k_y-k_0\sin\theta)}^2]\right).
\end{equation}
In radial coordinates $r=\sqrt{x^2+y^2}$, $k=\sqrt{k_x^2+k_y^2}$, the properly normalized wavefunctions read
\begin{equation}
\psi(r,t=0)=\frac{\rme^{-r^2/(2a^2)}\rme^{a^2 k_0^2/4}}
{a\sqrt{\pi\;\BesselI_0(a^2 k_0^2/2)}}\BesselJ_0(k_0 r),\quad
\hat{\psi}(k,t=0)=\frac{a\rme^{-a^2k^2/2-1/4a^2k_0^2}}{\sqrt{\pi\;\BesselI_0(a^2 k_0^2/2)}}\BesselI_0(a^2 k k_0),
\end{equation}
where $\BesselJ_0$ and $\BesselI_0$ denote Bessel functions \cite{Abramowitz1965a}.
The kinetic energy probability density
\begin{equation}\label{eq:momdis}
\frac{2\pi m}{\hbar^2}{\left|\hat{\psi}(k=\sqrt{2mE}/\hbar,t=0)\right|}^2
\end{equation}
is shown in Fig.~\ref{fig:mf_wp}. The choice of $a$ and $k_0$ is such, that the energy range under consideration is well covered. The wavepacket propagates via the DFFT method in $N=12000$ steps of $\Delta T=10^{-15}$~s. The grid consists of $300\times 300$ grid cells of width $2.67$~nm. The propagation takes about five minutes on a personal computer (2 GHz CPU).
Along the four edges of the grid we have placed a complex-valued potential of the form
\begin{equation}
V_{\rm abs}(r_i)=\rmi\;V_{0,\rm abs}\;\cosh^{-2}\left(\frac{{(r_i-r_{\rm edge})}^2}{d^2}\right),
\end{equation}
with  $V_{0,\rm abs}=25$~meV, $d=25$~nm. During the propagation we track for each grid point $r_{n,m}$ several quantities, including
\begin{equation}\label{eq:green}
G(\mbfr_{n,m},\mbfr';E)=\left[\sum_{j=1}^N\rme^{\rmi\,j\, \Delta t\,E/\hbar}\psi(\mbfr_{n,m},j\Delta t)\right] \frac{1}{\Delta(E)}
\end{equation}
for a set of energies $E$. If we want to obtain the Green function for a reasonable small region (i.e.\ around
the two constrictions), we can store $\psi(\mbfr_{n,m},j\Delta t)$ and obtain the energy-dependent Green function via the Laplace transform for any energy we pick. Thus a single run of the wavepacket yields $G(\mbfr_{n,m},\mbfr';E)$ for many energies and facilitates i.e.\ thermal averaging considerably.

In Fig.~\ref{fig:mf_jdensity} we display the microscopic current density originating from $\mbfr'$
\begin{equation}\label{eq:jden}
\mbfj(\mbfr_{n,m},\mbfr';E)=\frac{\hbar}{m}
\Im\left[G(\mbfr_{n,m},\mbfr';E)^*\nabla G(\mbfr_{n,m},\mbfr';E)\right]+\frac{e}{m}\AVP {|G(\mbfr_{n,m},\mbfr';E)|}^2
\end{equation}
throughout the whole device for $E=9$~meV. The obstacle around $\mbfr=(-350,140)$~nm deflects the current and lowers the transmission from the right constriction to the left one. Experimentally, it is possible to use a moveable charged tip as obstacle which is moved over the sample. High-resolution spatial information (10 nm) of the electronic current is obtained by recording the change in the transmitted current between the fixed constrictions as a function of tip position \cite{Aidala2007a}.
The resulting pictures reveal interference structures and flux bundles, along which the electronic current is flowing through the device. 

\begin{figure}[t]
\centerline {
\includegraphics[width=6in]{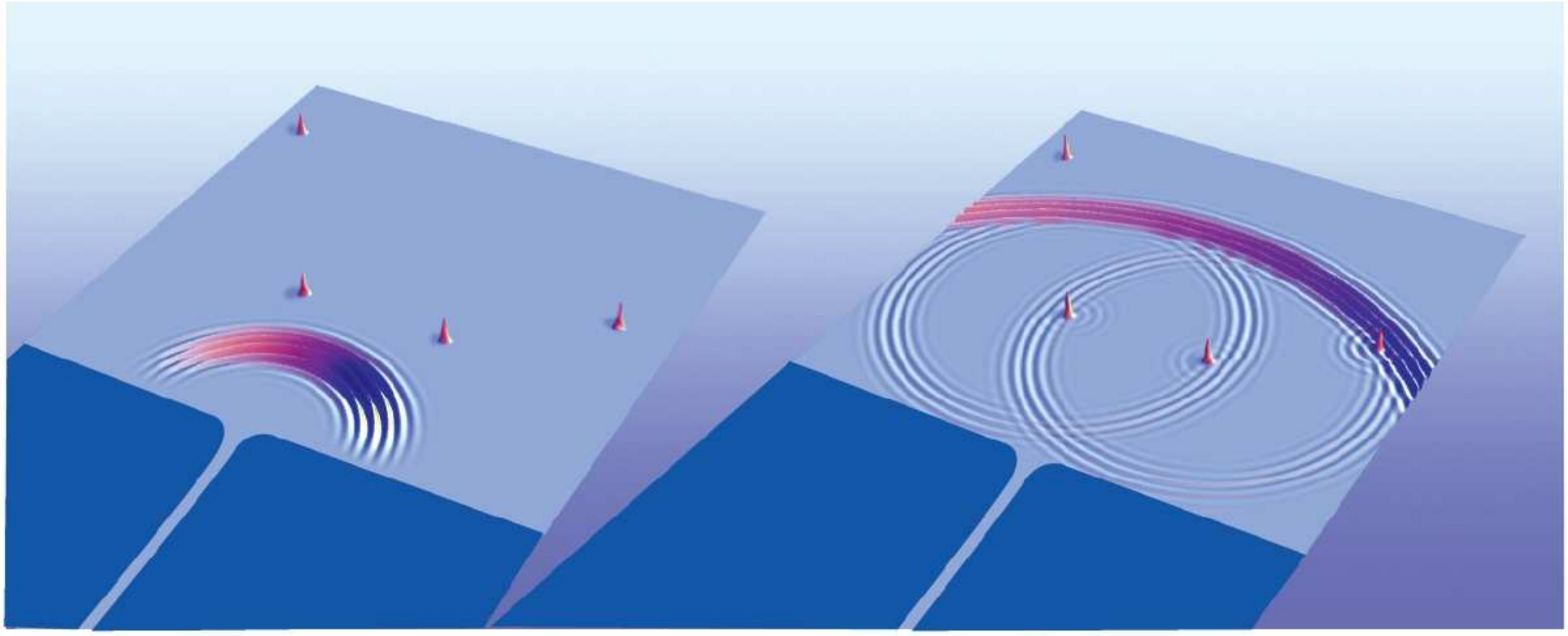}
}
\caption{A thermal wavepacket emerging from the quantum point contact (QPC) encounters two impurities which are approximately at the same radial distance from the QPC; the amplitude from these can interfere at the QPC (right); the amplitude scattered from impurities encountered later cannot interfere with the amplitude from the first two, since the arrival of that amplitude  will be  too late. }
\label{f2}
\end{figure}

\section{Wavepackets mimicking a thermal energy distribution}\label{sec:temp}

In order to obtain the energy-dependent Green function, we have discussed ways to largely remove the effect of wavepacket parameters, like the halfwidth $a$ of the initial wavepacket. However, it is also possible to retain the wavepacket properties and use them to model a certain occupation of initial energies. In the case of an adiabatic  quantum point contact (QPC) with one open transverse mode, a thermal ``smokering''  wavepacket with an energy distribution matching the derivative of the Fermi-Dirac distribution $-\partial f_{FD}({E'},E_F,T)/\partial E'$ replaces thermal averaging \cite{Heller2005a}. It is interesting that higher temperatures are represented by more confined wavepackets in position space, which reduces interference effects due to the shorter wave train. When considering different modes of backscattering which reduce the net flux through the QPC,  the question arises as to whether they can interfere. The answer is that they interfere only if the backscattered flux from the thermal wavepacket arrives at the QPC at the same time (see Fig.~\ref{f2}). This is a form of ``white light interferometry''.

The thermal wavepacket approach permits a powerful intuitive picture  and a simple estimation tool for understanding interference effects and electron choreography in small devices.  The basic scenario is as follows: Launched from just to the left of the QPC, a spatially narrow wavepacket emerges to the right, fanning out as part of an annulus, then suffering small angle scattering,  splitting up into pieces due to collisions with large and small objects, walls, etc.\ (Fig.~\ref{f2}). The pieces will have the same width in their direction of travel as the original wavepacket (unless the scattering is resonant and therefore time delaying).  As the pieces arrive at the original QPC (or another terminal) the rule is simple: separate pieces of the wavepacket must arrive at the same time  if they are to interfere.  (If the pieces exit to a lead with several modes open, they interfere mode by mode). If an object which the wavepacket encounters is moveable, we may follow the oscillation (in conductance) as the interference of the objects' phase changes from constructive to destructive and back to constructive as it is moved in certain directions.

\section{Conclusions}

We have obtained the energy-dependent Green function for general potentials via time-dependent wavepacket methods. The analytical solved cases allow to quantify and to minimize the error of the method due to the influence of wavepacket parameters. The wavepacket approach is especially useful for setups where periodic boundary conditions lead to non-physical results. The modelling of sources and sinks by complex potentials permits us to use very realistic potentials for devices and contacts and is at the same time efficient and fast. The time-dependent approach allows to postselect the energy (one or several) for which the energy-dependent Green function is needed. In mesoscopic physics, the wavepacket Green function describes transport in realistic nano device potentials and leads to new insights for the design and operation of future devices. A straightforward extension of the wavepacket methods allows one to include spin-orbit interactions by using two wavepackets corresponding to the two spin-components.

\subsection*{Acknowledment}

We appreciate helpful discussions with Katherine Aidala, Manfred Kleber, Peter Kramer, Michael Stopa, Robert Westervelt, and Michael Wimmer. The authors are grateful to the WE-Heraeus Foundation for the financial support to present the work at the 395th WE-Heraeus Seminar ``Time-dependent phenomena in Quantum Mechanics'' in Blaubeuren, Germany. TK is supported by the Emmy-Noether program of the DFG, grant Kr 2889/2-2.

\providecommand{\url}[1]{#1}

\end{document}